\documentstyle[floats,prd,aps,epsfig,eqsecnum,12pt]{revtex}

\makeatletter
\newbox\tempboxa
\newdimen\captionboxsubcount
\def\capsize#1{\captionboxsubcount=#1pt}
\newdimen\captionboxsub
\captionboxsub=\hsize \advance\captionboxsub by -\captionboxsubcount
\advance\captionboxsub by -\captionboxsubcount
\long
\def\@makecaption#1#2{
 \setbox\@tempboxa\hbox{#1 #2}
 \ifdim \wd\@tempboxa >\captionboxsub
\rightskip=\captionboxsubcount \leftskip=\captionboxsubcount #1 #2
\else \hbox to\hsize{\hfil\box\@tempboxa\hfil}
 \fi}
\makeatother
\capsize{30}

\begin{document}

\vfill
\begin{titlepage}
\begin{flushright}
\begin{minipage}{5cm}
\begin{flushleft}
\small
\baselineskip = 10pt
YCTP-P-04-00
\end{flushleft}
\end{minipage}
\end{flushright}

\begin{center}
\Large\bf
\centerline{Low Energy Theory for 2 flavors at High Density QCD}
\end{center}
\footnotesep = 12pt
\vfill
\begin{center}
\large \centerline{Roberto {\sc Casalbuoni}$^{a}$
\footnote{Electronic address: {\tt Roberto.Casalbuoni@fi.infn.it}}
\quad Zhiyong {\sc Duan}$^b$ \footnote{Electronic address: {\tt
zhiyong.duan@yale.edu}} \quad Francesco {\sc Sannino}$^b$
\footnote{ Electronic address: {\tt francesco.sannino@yale.edu}}}
\vskip .5cm $^{a}${\it  Dipartimento Di Fisica, Univ. di Firenze
and I.N.F.N Sezione di Firenze,  I-50125, Italia.} \vskip .5cm
$^b${\it Department of Physics, Yale Univ., New Haven,  CT
06520-8120, USA.}
\end{center}
\vfill
\begin{center}
\bf Abstract
\end{center}
\begin{abstract}
\baselineskip = 17pt We construct the effective Lagrangian
describing the low energy excitations for Quantum Chromodynamics
with two flavors at high density. The non-linear realization
framework is employed to properly construct the low energy
effective theory. The light degrees of freedom, as required by
't~Hooft anomaly conditions, contain massless fermions which we
properly include in the effective Lagrangian. We also provide a
discussion of the linearly realized Lagrangian.
\end{abstract}
\begin{flushleft}
\footnotesize
\end{flushleft}
\vfill
\end{titlepage}

\section{Introduction}

\label{uno}

Recently quark matter at very high density has attracted a great
flurry of interest \cite{ARW_98,RSSV_98,ARW_98b,REV,SW_98b}. In
this region, quark matter is expected to behave as a color
superconductor \cite{ARW_98,RSSV_98}. Possible phenomenological
applications are associated with the description of neutron star
interiors, neutron star collisions and the physics near the core of
collapsing stars. A better understanding of highly squeezed nuclear
matter might also shed some light on nuclear matter at low density,
i.e. densities close to ordinary nuclear matter where some models
already exist. For example, in Ref.~\cite{HSSW} a rather complete
soliton model at low density is constructed containing
vector-bosons, along with the Goldstone bosons.

In a superconductive phase, the color symmetry is spontaneously
broken and a hierarchy of scales, for given chemical potential, is
generated. Indicating with $g$, the underlying coupling constant,
the relevant scales are: the chemical potential $\mu $ itself, the
dynamically generated gluon mass $m_{gluon}\sim g\mu $ and the gap
parameter $\Delta \sim \frac{\mu }{g^{5}}e^{-\frac{\alpha }{g}}$
with $\alpha$ a calculable constant. Since for high $\mu$ the
coupling constant $g$ (evaluated at the fixed scale $\mu $) is $\ll
1$ , we have:
\begin{equation}
\Delta \ll m_{gluon}\ll \mu \ .
\end{equation}
Massless excitations dominate physical processes at very low energy
with respect to the energy gap ($\Delta$). Their spectrum is
intimately related to the underlying global symmetries and the way
they are realized at low energies. Indeed when the dynamics is such
that a continuous global symmetry is spontaneously broken, a
Goldstone boson appears in order to compensate for the breaking.
Massless excitations obey low energy theorems governing their
interactions which can be usefully encoded in effective
Lagrangians. A well known example, in the region of cold and
non-dense QCD, is the effective Lagrangian for pions and kaons.
These Lagrangians are seen to describe well the QCD low energy
phenomenology \cite{effective Lagrangians}.

Another set of relevant constraints is provided by quantum
anomalies. At zero density and temperature, 't~Hooft \cite{thooft}
argued that the underlying continuous global anomalies have to be
matched in a given low energy phase by a set of massless fermions
associated with the intact global symmetries and a set of massless
Goldstone bosons associated with the broken ones. The low energy
fermions (composite or elementary) contribute via triangle diagrams
while for the Goldstones a Wess-Zumino term should be added to
correctly implement the associated global anomalies.

In \cite{S}, the 't~Hooft constraints were seen to hold for QCD at
finite density. In particular, it was shown, by reviewing the
dynamically favored phases for $N_{f}=2,3$ at high density, that
the low energy spectrum displayes the correct quantum numbers to
saturate the 't~Hooft global anomalies. It was also observed that
QCD at finite density can be envisioned, from a global symmetry and
anomaly point of view, as a chiral gauge theory \cite{ball,ADS} for
which at least part of the matter field content is in complex
representations of the gauge group. Indeed an important distinction
from zero density, vector-like theories is that these theories,
when strongly coupled, can exist in the Higgs phase by dynamically
breaking their own gauge symmetries\cite{ball,ADS}. This is also
the striking feature of the superconductive phase allowed for QCD
at high density. In fact at finite density, vector-like symmetries
are no longer protected against spontaneous breaking by the
Vafa-Witten theorem \cite{VW}.

In this paper we build the effective theory describing the low
energy excitations for Quantum Chromodynamics with two flavors at
high density. The non linear realization framework \cite{CWZ} is
used to properly construct the low energy theory. The light degrees
of freedom, as required by 't~Hooft anomaly conditions, contains
massless fermions which we properly include in the effective
Lagrangian. We prove that the non linearly realized low energy
effective theory, in general, contains at the lowest order in a
derivative expansion for the Goldstones (i.e. two derivative) two
independent terms which are responsible for distinct contributions
to the gluon masses \cite{R} when gauging the color degrees of
freedom.

We finally provide a discussion of the linearly realized Lagrangian. In this
framework we see, as in reference \cite{R,ssh}, that the linearly realized
effective Lagrangian at the dimension four level actually predicts the
squared gluon mass ratio between the eighth gluon and the other massive ones
to be $4/3$. This result does not agree with the underlying calculations
\cite{R}. In order to resolve the issue, in Ref.~\cite{R,ssh} it was pointed
out that one can include in the linear Lagrangian another two derivative
term possessing mass dimension 6. This result shows the failure of the
standard naive dimensional counting arguments for suppressing higher
dimensional terms for the effective linearly realized effective Lagrangian.
Interestingly, in the non linear realization case all the terms needed to
correctly describe the gluon masses arises at the same derivative order.
This fact suggests that non linear realization description leads to a
consistent counting scheme.

In Section \ref{General} we summarize some general features of 2 flavor QCD.
The general non linear formalism for the spontaneous breaking of color is
presented in Section \ref{Low}. We implement the breaking of the baryon
number in Section \ref{abelian}. Then, in medium, fermions are introduced in
Section \ref{fermions} together with the summary of the low energy theory
and its extension to the broken Lorentz group case. A linearly realized
Lagrangian is explored in Section {\ref{linear}}. In Section {\ref
{conclusions}} we conclude.

\section{General features of 2 flavor QCD at High Density}
\label{General}

The underlying gauge group is $SU(3)$ while the quantum flavor group is
\begin{equation}
SU_{L}(2)\times SU_{R}(2)\times U_{V}(1)\ ,
\end{equation}
and the classical $U_{A}(1)$ symmetry is destroyed at the quantum
level by the Adler-Bell-Jackiw anomaly. We indicate with
$q_{Lc,i;\alpha }$ the two component left spinor where $\alpha
=1,2$ is the spin index, $c=1,2,3$ is the color index while $i=1,2$
represents the flavor. ${q_{R}}_{{c,i}}^{\dot{\alpha}}$ is the two
component right spinor. We summarize the transformation properties
in the following table:
\begin{equation}
\begin{tabular}{ccccc}
& $[SU(3)]$ & $SU_{L}(2)$ & $SU_{R}(2)$ & $U_{V}(1)$ \\
&  &  &  &  \\
$q_{L}$ &
\begin{tabular}{|c|}
\hline
$\ \ $ \\ \hline
\end{tabular}
&
\begin{tabular}{|c|}
\hline
$\ \ $ \\ \hline
\end{tabular}
& $1$ & $1$ \\
&  &  &  &  \\
${q_{R}}$ & $
\begin{tabular}{|c|}
\hline
$\ \ $ \\ \hline
\end{tabular}
$ & $1$ & $
\begin{tabular}{|c|}
\hline
$\ \ $ \\ \hline
\end{tabular}
$ & $1$%
\end{tabular}
\end{equation}
where $SU(3)$ is the gauge group, indicated by square bracket. The
theory is subject to the following global anomalies:
\begin{equation}
SU_{L/R}(2)^{2}U_{V}(1)\propto \pm 3\ ,
\end{equation}
where we have chosen a convention for the flavor generators in which the
quadratic anomaly factor is $1$.

At zero density we have two possible phases compatible with the
anomaly conditions. One is the ordinary Goldstone phase associated
with the spontaneous breaking of the underlying global symmetry to
$SU_{V}(2)\times U_{V}(1)$. The other is the Wigner-Weyl phase
where, assuming confinement, the global symmetry at low energy is
intact and the needed massless spectrum consists of massless
baryons. The Goldstone phase, associated with a non vanishing
$\bar{q}_{L}q_{R}$ condensate, is the one observed in nature. This
fact supports a new idea presented in Ref.~\cite{ADS}. There it is
suggested that, for an asymptotically free theory, among multiple
infrared phases allowed by 't~Hooft anomaly conditions, the one
which minimizes the entropy at the approach to freeze-out is
preferred.

What happens when we squeeze nuclear matter? At very low baryon density
compared to a fixed intrinsic scale of the theory $\Lambda $, it is
reasonable to expect that the Goldstone phase persists. On the other hand at
very large densities, it is seen, via dynamical calculations \cite
{ARW_98,RSSV_98}, that the ordinary Goldstone phase is no longer favored
compared with a superconductive one associated with the following type of
quark condensates:
\begin{equation}
\epsilon ^{\alpha \beta }\epsilon ^{abc}\epsilon ^{ij}<q_{Lb,i;\alpha
}q_{Lc,j;\beta }>\ ,\qquad \epsilon _{\dot{\alpha}\dot{\beta}}\epsilon
^{abc}\epsilon ^{ij}<q_{Rb,i}^{\dot{\alpha}}q_{Rc,j}^{\dot{\beta}}>\ .
\label{condensation}
\end{equation}
We can now introduce two scalar fields which play the role of order
parameters:
\begin{equation}
{L^{\dagger }}^{a}\sim \epsilon ^{abc}\epsilon ^{ij}q_{Lb,i}^{\alpha
}q_{Lc,j;\alpha }\ ,\qquad {R^{\dagger }}^{a}\sim -\epsilon ^{abc}\epsilon
^{ij}q_{Rb,i;\dot{\alpha}}q_{Rc,j}^{\dot{\alpha}}\ .
\end{equation}
Under a parity transformation
\begin{equation}
L_{a}\leftrightarrow R_{a}.
\end{equation}
If parity is not broken spontaneously, we have
\begin{equation}
\left\langle L_{a}\right\rangle =\left\langle R_{a}\right\rangle =v\delta
_{a}^{3}\ ,  \label{vev}
\end{equation}
where we chose the condensate to be in the 3rd direction of color.
This condensate is not allowed at zero density by the Vafa-Witten
theorem. The order parameters are singlets under the
$SU_{L}(2)\times SU_{R}(2)$ flavor transformations while possessing
baryon charge $2$.

The vev breaks the gauge symmetry while leaving intact the
following group:
\begin{equation}
\left[ SU(2)\right] \times SU_{L}(2)\times SU_{R}(2)\times \widetilde{U}_{V}(1)\ ,
\end{equation}
where $\left[ SU(2)\right] $ is the unbroken part of the gauge
group. The $\widetilde{U}_{V}(1)$ generator is the following linear
combination of the previous $U_{V}(1)$ generator $Q={\rm
diag}(1,1,1)$ and the broken diagonal generator of the $SU(3)$
gauge group $T^{8}=\frac{1}{2\sqrt{3}}\,{\rm diag}(1,1,-2)$
\begin{equation}
\widetilde{S}=\frac{1}{3\sqrt{2}}\left[Q-2\sqrt{3}T^{8}\right]\ .  \label{residue}
\end{equation}
The $\widetilde{S}$ charge of the quarks with color $1$ and $2$ is
zero.

The superconductive phase for $N_f=2$ possesses the same global
symmetry group of the confined Wigner-Weyl phase. This remarkable
feature when considering chiral gauge theories at zero density
(where the superconductive phase is now a Higgs phase) is referred
as complementarity. This idea was introduced in Ref.~\cite{rds}
where it was conjectured that any Higgs phase can be described in
terms of confined degrees of freedom and vice versa.

It is convenient to introduce the fields:
\begin{equation}
V_{a}=\frac{L_{a}+R_{a}}{\sqrt{2}}\ ,\qquad A=\frac{L_{a}-R_{a}}{\sqrt{2}}\ .
\end{equation}
On the vacuum, via (\ref{vev}), we have:
\begin{equation}
\left\langle V_{a}\right\rangle =\sqrt{2}v\delta _{a}^{3}\qquad \left\langle
A_{a}\right\rangle =0\ .  \label{svev}
\end{equation}
The massless excitations are associated with the fluctuations
around the vacuum expectation value for $V_{a}$ while the fields
described by $A_{a}$ are massive. In Ref.~\cite{MSW} it is argued,
based on a dynamical calculation, that at very high density the
$A_{a}$ fields might be very light and possibly relevant for the
low energy phenomenology. However we will focus mainly on the truly
massless excitations since low energy theorems are valid only for
these fields. Low energy theorems are, in general, efficiently
encoded in a non linear realization framework which we will soon
construct.

\section{Low Energy Effective Theory without Baryon number}

\label{Low}

In order to write an effective Lagrangian for 2 flavor QCD, one could, at
least in principle, proceed in different ways. For example one could add a
mass term for the strange quark in the $3$ flavor Lagrangian. However, since
the theory is not supersymmetric, an exact decoupling procedure is not known%
\footnote{An attempt to generalize to QCD the Seiberg decoupling
procedure at the effective Lagrangian level is provided in
\cite{HSS}.} at effective Lagrangian level. A general way, which we
will explore here, to directly construct the 2 flavor effective
Lagrangian makes use of the non linear realization methods
\cite{CWZ}. The latter have been already successfully employed for
the 3 flavor case at high density in \cite{CG}.

Following reference \cite{CG}, we now construct the non linearly realized
effective Lagrangian containing the diquark degrees of freedom coupled to
gluons.

We postpone the breaking of the baryon number and its consequences
to the next section. The group of transformations is $G=SU(3)$,
while the stability group leaving the vacuum invariant is the
proper subgroup $H=SU(2)\subset G$. The color generators $T^{m}$ of
$SU(3)$, with $m=1,\ldots ,8,$ obey the normalization condition
${\rm Tr}\left[ T^{m}T^{n}\right] =\frac{1}{2}\delta ^{mn}$. We
divide the generators $\{T\}$ into two classes, calling the
generators of $H$ $\{S^{a}=T^{a}\}$, with $a=1,\ldots ,3$ and the
broken generators $\{X^{i}=T^{i+3}\}$ with $i=1,\ldots ,5$. It is
worth noticing that the quotient space $G/H$, in this case, is not
a symmetric space. By symmetric space we mean that, if $X$ and $S$
represent arbitrary linear combinations of the broken and unbroken
generators, their commutators should satisfy the restriction:
\begin{equation}
\left[ X,X\right] =iS\ .  \label{ss}
\end{equation}
It is easy to verify that the previous condition is not obeyed. Clearly, our
generators should always satisfy the trivial conditions
\begin{equation}
\left[ S,S\right] =iS\ ,\qquad \left[ X,S\right] =iX\ ,
\end{equation}
expressing the fact that the Lie algebra of $H$ closes and that the $\{X\}$
form a representation of $H$.

The coset space $G/H$ is parameterized by the group elements
\begin{equation}
{\cal V}=\exp (i\xi ^{i}X^{i})\ ,
\end{equation}
with $\xi ^{i}=\Pi ^{i}/v$ describing the Goldstone fields and are
coordinates of the space $G/H$. ${\cal V}$ transforms non linearly under a
color transformation, i.e.:
\begin{equation}
{\cal V}(\xi )\rightarrow g{\cal V}(\xi )h^{\dagger }(\xi ,g)\
,\label{nl}
\end{equation}
where $g\in G$ and $h\in H$. It is convenient to define the
hermitian (algebra valued) Maurer-Cartan one-form
\begin{equation}
\omega_\mu=i {\cal V}^\dagger D_\mu {\cal V} \ ,
\end{equation}
with $D_\mu$ the covariant derivative with respect to $G=SU(3)$
\begin{equation}
D_\mu {\cal V}=(\partial_\mu - iG_\mu) {\cal V} \ ,
\end{equation}
and $G_{\mu}^m T^m$ the gluon fields. Since ${\cal V}$ transforms with
respect to $G$ as in Eq.~(\ref{nl}) it follows that
\begin{equation}
\omega_\mu\to h(\xi,g) \omega_{\mu} h^\dagger(\xi,g)+h(\xi,g)\partial{_\mu}
h^\dagger(\xi,g) \ .
\end{equation}
We decompose $\omega_\mu$ into the part parallel to $H$
\begin{equation}
\omega_\mu^{\parallel}= 2 S^a {\rm Tr}\left[S^a \omega_{\mu} \right]\ ,
\qquad \in {\rm Lie}~H \ ,
\end{equation}
and into the perpendicular part
\begin{equation}
\omega_\mu^{\perp}= 2 X^i {\rm Tr}\left[X^i \omega_{\mu} \right]\ ,\qquad
\in {\rm Lie}~G-{\rm Lie}~H \ .
\end{equation}
Summation over repeated indices is assumed.

This yields the following transformation properties:
\begin{equation}
\omega _{\mu }^{\parallel }\rightarrow h(\xi ,g)\omega _{\mu }^{\parallel
}h^{\dagger }(\xi ,g)+h(\xi ,g)\partial _{\mu }h^{\dagger }(\xi ,g)\ ,
\end{equation}
and
\begin{equation}
\omega _{\mu }^{\perp }\rightarrow h(\xi ,g)\omega _{\mu }^{\perp
}h^{\dagger }(\xi ,g)\ .
\end{equation}
Then, it turns out that the most general invariant of second order in the
derivatives is
\begin{equation}
L=v^{2}{\rm Tr}\left[ \omega _{\mu }^{\perp }\omega ^{\mu \perp
}\right] =-2\,v^{2}\,{\rm Tr}\left[ X^{i}{\cal V}^{\dagger }D_{\mu
}{\cal V}\right] \,{\rm Tr}\left[ X^{i}{\cal V}^{\dagger }D^{\mu
}{\cal V}\right] \ .
\end{equation}
We can immediately see, by adopting the unitary gauge (corresponding to $\xi
\rightarrow 0$), that the previous term provides a mass term for the gluons
associated with the 5 coset generators, whereas the gluons
$G^{1,2,3}$ remain massless. Indeed the mass term reads:
\begin{equation}
L=2v^{2}{\rm Tr}\left[ X^{i}G_{\mu }\right] {\rm Tr}\left[
X^{i}G^{\mu }\right] =\frac{v^{2}}{2}G_{\mu }^{i}{G^{i{\mu }}}\ .
\end{equation}
The fact that we find the same mass for all of the gluons of the broken
subgroup is not surprising, since we have treated them equally. In the next
section we address the problem of the baryon number and show how, in the non
linear framework, this naturally leads to a distinct mass for the eighth
gluon.

The massless gluons of the subgroup $H$ confine, leaving, as we shall see,
only some of the quarks as massless excitations at low energies.

\section{The Abelian Group $U_{V}(1)$}

\label{abelian}

The previous discussion is not yet complete, since we have omitted the
breaking of the baryon number. Indeed the group of transformations should be
$G=SU(3)\times U_{V}(1)$, while the unbroken subgroup is $H=SU(2)\times
\widetilde{U}_{V}(1)$.

The non linear transformations are now:
\begin{equation}
{\cal V}(\xi )\rightarrow u_{V}\,g{\cal V}(\xi )h^{\dagger }(\xi
,g,u)h_{\widetilde{V}}^{\dagger }(\xi ,g,u)\ ,  \label{nl2}
\end{equation}
\begin{equation}
u_{V}\in U_{V}(1)\ ,\quad g\in SU(3)\ ,\quad h(\xi ,g,u)\in SU(2)\
,\quad h_{\widetilde{V}}(\xi ,g,u)\in \widetilde{U}_{V}(1)\ .
\end{equation}
The $U_{V}(1)$ charge is $Q={\rm diag}(1,1,1)$. The generator, for
$\widetilde{U}_{V}(1)$, leaving the vev invariant is
\begin{equation}
\widetilde{S}=\frac{1}{3\sqrt{2}}\left[ Q-2\sqrt{3}T^{8}\right] \
= {\rm diag}(0,0,{\sqrt{2}}/{2}),  \label{tildeS}
\end{equation}
where we have chosen to normalize it according to Tr$\left[
\widetilde{S}^{2}\right] =1/2$. The coset space $G/H$ is
parameterized, as before, by the group elements
\begin{equation}
{\cal V}=\exp (i\xi ^{i}X^{i})\ ,
\end{equation}
but now we identify the Goldstone bosons as:
\begin{eqnarray}
\xi
^{i}&=&\frac{\Pi^{i}}{v} \quad i=1,2,3,4 \ , \\
\xi^5 &=& \frac{\Pi^5}{\widetilde{v}} \ .
\label{tildev}
\end{eqnarray}
We still have 5 generators $\{X^{i}\}$ which belong to the coset
space $G/H$, but one of them needs to be modified. While
$X^{1,2,3,4}$ are still identified, respectively, with
$T^{4,5,6,7}$, the coset generator which replaces $X^{5}=T^{8}$ of
the previous section is now
\begin{equation}
X^{5}=\frac{1}{3}\left[ Q+\sqrt{3}T^{8}\right] ={\rm
diag}(\frac{1}{2},\frac{1}{2},0).
\end{equation}
We used the orthogonality condition
\begin{equation}
{\rm Tr}\left[ XS\right] =0\ ,
\end{equation}
to construct $X^{5}$ with Tr$\left[ X^{5}X^{5}\right] =\frac{1}{2}$. Notice
that $X^{5}$ is no longer traceless (i.e. ${\rm Tr}\left[ X^{5}\right] =1$).

It is amusing to note that ${\cal V}$ transforms on the left as an ordinary
quark with respect to color and $U_V(1)$ transformations. This will turn to
be important when adding the fermions.

As for ordinary QCD, we can construct a field transforming linearly by using
only non linearly realized fields via:
\begin{equation}
V_{a}=\frac{v}{\sqrt{2}}\,\epsilon _{abc}\,{\cal V}_{i}^{b}{\cal V}
_{j}^{c}\,\epsilon ^{ij3}\ .  \label{lvnl}
\end{equation}
To prove that the previous expression transforms linearly we recall
that the $H$ subgroup acts on the right of ${\cal V}$. So under a
general non linear transformation we have:
\begin{equation}
\epsilon _{abc}\,{\cal V}_{i}^{b}{\cal V} _{j}^{c}\,\epsilon
^{ij3} \rightarrow u_V^2 \epsilon _{abc} g_d^b g_e^c {\cal
V}_{k}^{d}{\cal V} _{l}^{e} {h^{\dagger}}^k_i {h^{\dagger}}^l_j
{h^{\dagger}_{\widetilde{V}}} ^2\,\epsilon ^{ij3} \rightarrow
u_V^2 \epsilon _{abc} g_d^b g_e^c {\cal V}_{k}^{d}{\cal V}
_{l}^{e}\, \epsilon ^{kl3} \ .
\end{equation}
Indeed due to Eq.~(\ref{tildeS}) we have that $h_{\widetilde{V}}$
acting on the color indices $1$ and $2$ is equivalent to the
identity. We also have
\begin{equation}
{h^{\dagger}}^k_i {h^{\dagger}}^l_j \,\epsilon ^{ij3} =\,\epsilon^{kl3} \ ,
\end{equation}
since $h$ is an $SU(2)$ matrix and we have conveniently chosen the
indices. Therefore, $V_a$ transforms linearly. Using the relation:
\begin{equation}
{\rm det}{\cal V}\equiv\frac{1}{3 !}\, \epsilon
_{abc}\,\epsilon^{kij}\,{\cal V}_{k}^{a} {\cal V}_{i}^{b}{\cal V}
_{j}^{c}=\exp\left[i\frac{\Pi^5}{\widetilde{v}} \frac{{\rm Tr}\left[Q\right]}
{3}\right] =
 \exp\left[i\frac{\Pi^5}{\widetilde{v}} \right]\ ,
\end{equation}
we can rewrite Eq.~(\ref{lvnl}) in the following form:
\begin{equation}
V_{a}=\sqrt{2}\, v \, e^{i \frac{\Pi^5}{\widetilde{v}}}\, {{\cal
V}^{-1}}^3_a
\ .
\end{equation}
The latter expression corresponds to the polar decomposition of
the vector $V_a$ with the massive scalar field frozen to its vev
value. Since the field $\Pi^5$ is associated with the $X^5$
generator, which is a linear combination of an abelian and of a
non abelian $T^8$ generator, ${\rm det}{\cal V}$ does not
parameterize an independent field.

Expanding ${\cal V}$ up to first order in the Goldstones leads to:
\begin{equation}
V_{a}=\sqrt{2}\,v\,\left[1 + i \xi^5\left(\frac{2}{3}Q
- \frac{1}{\sqrt{3}}T^8\right)-i\sum_{i=1}^4\xi^i X^i\right]_{a}^{c}\delta
_{c}^{3}+v\,{\cal O}\left(\xi ^{2}\right) \ .
\end{equation}
This new field explicitly describes the vev properties. {}First of all, when
considering the limit $\xi \rightarrow 0$ we recover the correct vacuum,
i.e.
\begin{equation}
V_{a}=\sqrt{2}v\delta _{a}^{3}\ .
\end{equation}
Furthermore, as expected, $V_{a}$ transforms under the underlying
gauge transformations as a diquark.

Equation (\ref{lvnl}) mimics what happens in ordinary QCD. There
the vev transformations are encoded in the linearly transforming
matrix $U\rightarrow g_{L}Ug_{R}^{\dagger }$ while the non linearly
transforming field is simply $\sqrt{U}\rightarrow
g_{L}\sqrt{U}k^{\dagger }(\sqrt{U}
,g_{L},g_{R})=k(\sqrt{U},g_{L},g_{R})\sqrt{U}g_{R}^{\dagger }$,
with $k(\sqrt{U},g_{L},g_{R})\in SU_{V}(3)$.

If we insist in gauging the color transformations the general
discussion in the previous section remains unchanged. The
substantial difference is that now the most generic two derivative
kinetic Lagrangian
\begin{equation}
L=v^{2}a_{1}{\rm Tr}\left[ \,\omega _{\mu }^{\perp }\omega ^{\mu
\perp }\, \right] +v^{2}a_{2}{\rm Tr}\left[ \,\omega _{\mu
}^{\perp }\,\right] {\rm Tr} \left[ \,\omega ^{\mu \perp }\,\right]
\ ,  \label{dt}
\end{equation}
acquires a new term. The presence of a double trace term is due to the
absence of the traceless condition for the broken generator $X^{5}$.

The kinetic term for the Goldstones is:
\begin{equation}
L_{kin}= \frac{a_1}{2} \sum_{i=1}^4 \partial_{\mu} \Pi^i
\partial^{\mu} \Pi^i + \frac{v^2}{\widetilde{v}^2}
\left(\frac{a_1+2a_2}{2}\right)\partial_{\mu} \Pi^5 \partial^{\mu} \Pi^5 \ .
\end{equation}
Normalizing the kinetic term we have
\begin{equation}
a_1=1 \ , \qquad a_2=\frac{1}{2}
\left[\frac{\widetilde{v}^2}{{v}^2} -a_1\right] \ .
\label{rel}
\end{equation}

The square gluon mass ratio, in the unitary gauge (i.e. $\xi
\rightarrow 0$), between the eighth gluon and any other massive one now
reads:
\begin{equation}
\frac{m_{8}^{2}}{m_{i}^{2}}=\frac{1}{3}\left[ 1+2\frac{a_{2}}{a_1}\right]=
\frac{1}{3}\frac{\widetilde{v}^2}{{v}^2}
\ ,
\end{equation}
with $i=4,5,6$ or $7$ and in the last step we use Eq.~(\ref{rel}).
Using perturbation theory at one loop \cite{R} $a_{2}\approx 1/2$
(i.e. $\widetilde{v}\approx\sqrt{2}\,v$). Also, as we shall see, in
the linearly realized lagrangian one can write \cite{ssh} two
two-derivative terms in the effective Lagrangian. However one of
them appears in the Lagrangian as a higher order in mass dimension,
leading to a not straightforward systematic expansion.

\section{In Medium Fermions}

\label{fermions}

Now we turn our attention to the Fermi fields. First of all, let us
introduce the following field:
\begin{equation}
\widetilde{\psi}={\cal V}^{\dagger }\psi \ ,  \label{mq}
\end{equation}
which under the action of the group $G=SU(3)\times U_V(1)$ transforms as
\begin{equation}
\widetilde{\psi}\rightarrow h_{\widetilde{V}}(\xi,g,u)h(\xi
,g,u)\widetilde{\psi}\ .
\end{equation}
This construction allows us to easily switch to the non linear
representations and it is often encountered in the Heavy Quark Effective
formalism \cite{HQ}.

The fermion field in (\ref{mq}) can be thought as a, in medium,
quark surrounded by a Goldstone cloud. Remarkably, as also required
by 't~Hooft anomaly conditions \cite{S}, the newly defined (in
medium) fermion possesses the unbroken $\widetilde{U}_V(1)$ charge.
The cloud (effectively included in ${\cal V}$) correctly screens
the baryonic charge.

In the new variables, one can construct two independent invariant terms
\begin{equation}
L_{1}=\overline{\widetilde{\psi }}i\gamma ^{\mu }(\partial _{\mu
}-i\omega _{\mu }^{\parallel })\widetilde{\psi }\ ,\qquad
L_{2}=\overline{\widetilde{\psi }}\gamma ^{\mu }\omega _{\mu
}^{\perp }\widetilde{\psi }\ .
\end{equation}
One can easily verify that the tree level lagrangian term for the quarks
\begin{equation}
L_{{\rm {Tree}}}=\bar{\psi}\gamma ^{\mu }(\partial _{\mu }-iG_{\mu })\psi \ ,
\end{equation}
corresponds to the linear combination of
\begin{equation}
L_{{\rm {Tree}}}=L_{1}+L_{2}\ .
\end{equation}

Another invariant that we can add to the effective Lagrangian corresponds to
the following Majorana mass term:
\begin{equation}
L_{m}=m_{M}\overline{\tilde{\psi}^{C}}_{i}\gamma ^{5}(iT^{2})\tilde{\psi}%
_{j}\epsilon ^{ij}+{\rm h.c.}\   \label{mmass}
\end{equation}
with $\tilde{\psi}^{C}=i\gamma ^{2}\tilde{\psi}^{\ast }$, where $i,j=1,2$
are flavor indices. The invariance under a $h(\xi ,g,u)\in SU(2)$
transformation is insured by the relation
\begin{equation}
(S^{a})^{T}T^{2}=-T^{2}S^{a},
\end{equation}
where
\begin{equation}
T^{2}=S^{2}=\frac{1}{2}\left(
\begin{array}{ll}
\sigma ^{2} & 0 \\
0 & 0
\end{array}
\right) \ .
\end{equation}
The invariance under a $\tilde{U}_{V}\left( 1\right)$ transformation is due
to the fact that its generator $\tilde{S}={\rm diag}(0,0,\frac{\sqrt{2}}{2})$
is such that $(\tilde{S})^{T}T^{2}=-T^{2}\tilde{S}=0$.

We stress that this Lagrangian term respects the underlying color
transformations as well as the global transformations
$SU_{L}(2)\times SU_{R}(2)\times U_{V}\left(1\right) $. The
breaking of the underlying symmetries manifests itself only when we
evaluate this Lagrangian on the vev (i.e. $\xi \rightarrow 0$). To
see this one can rewrite the previous terms as an explicit function
of ${\cal V}$. Parity is also enforced. This term does not produce
a mass term for the quarks with color index $3$ while yielding a
Majorana mass term for fermions with color index $1$ and $2$.

The latter Lagrangian term is similar to the meson-fermion
interaction we can write in standard QCD (of the type $\bar{q}Uq$
which yields the constituent Dirac quark masses when evaluated on
the vev).

At this point, for reader's convenience, we summarize the total non linearly
realized effective Lagrangian describing in medium fermions, gluons and
their self interactions, up to two derivatives,
\begin{eqnarray}
{\cal L}=~ &&v^{2}a_{1}{\rm Tr}\left[ \,\omega _{\mu }^{\perp
}\omega ^{\mu \perp }\,\right] +v^{2}a_{2}{\rm Tr}\left[ \,\omega
_{\mu }^{\perp }\,\right] {\rm Tr}\left[ \,\omega ^{\mu \perp
}\,\right]  \nonumber \\ &+&b_{1}\overline{\widetilde{\psi
}}i\gamma ^{\mu }(\partial _{\mu }-i\omega _{\mu }^{\parallel
})\widetilde{\psi }+b_{2}\overline{\widetilde{\psi }} \gamma
^{\mu }\omega _{\mu }^{\perp }\widetilde{\psi }  \nonumber
\\ &+&m_{M}\overline{\tilde{\psi}^{C}}\gamma
^{5}(iT^{2})\tilde{\psi}+{\rm h.c.} \ .
\end{eqnarray}
Here $a_{1},~a_{2},~b_{1}$ and $b_{2}$ are real coefficients while
$m_{M}$ is complex and we omit flavor indices. At the tree level in
the underlying theory we have $b_{1}=b_{2}=1 $. The massless
degrees of freedom are the in medium fermions
$\tilde{\psi}_{a=3,i}$ which possess the correct quantum numbers
prescribed by 't~Hooft anomaly conditions \cite{S}. As already
mentioned, the $SU(2)$ color subgroup remains unbroken and the
associated 3 gluons are expected to confine again. To the previous
general effective Lagrangian we should also add the gluon kinetic
term.

In writing the effective low energy theory we have not yet considered the
breaking of Lorentz invariance at finite density. Following Ref.~\cite{CG}
we impose invariance only under the $O(3)$ subgroup of the Lorentz
transformations. This amounts to have different coefficients for the
temporal and spatial indices of the Lagrangian which now becomes:
\begin{eqnarray}
{\cal L}=~ &&v^{2}a_{1}{\rm Tr}\left[ \,\omega _{0}^{\perp }\omega
_{0}^{\perp }-{\alpha }_{1}\vec{\omega} ^{\perp }\vec{\omega
}^{\perp }\, \right] +v^{2}a_{2}\left[ {\rm Tr}\left[ \,\omega
_{0}^{\perp }\,\right] {\rm Tr} \left[ \,\omega _{0}^{\perp
}\,\right] -{\alpha }_{2}{\rm Tr}\left[ \,\vec{\omega} ^{\perp
}\,\right] {\rm Tr}\left[ \,\vec{\omega} ^{\perp }\, \right]
\right]  \nonumber \\ &+&b_{1}\overline{\widetilde{\psi }}i\left[
\gamma ^{0}(\partial _{0}- i\omega _{0}^{\parallel })+\beta
_{1}\vec{\gamma}\cdot\left(\vec{\nabla} - i\vec{\omega}
^{\parallel}\right) \right] \widetilde{\psi
}+b_{2}\overline{\widetilde{ \psi }}\left[ \gamma ^{0}\omega
_{0}^{\perp }+\beta _{2}\vec{\gamma } \cdot \vec{\omega}^{\perp
}\right] \widetilde{\psi }  \nonumber \\
&+&m_{M}\overline{\tilde{\psi}^{C}}\gamma
^{5}(iT^{2})\tilde{\psi}+{\rm h.c.} \ ,
\label{final}
\end{eqnarray}
where the new coefficients $\alpha $s and $\beta $s encode the
effective breaking of Lorentz invariance. To the previous
Lagrangian we can still add the chemical potential type of term
$\displaystyle{\overline{\widetilde{\psi}}\gamma^0\widetilde{\psi}}$.

Clearly, in future, it would be valuable to compute the
coefficients of the effective Lagrangian in Eq.~(\ref{final}) at
asymptotically high densities. We notice that, due to the
invariance under the global $SU_L(2)\times SU_R(2)$ symmetry group,
no mass term arises for the third colored quarks and that any
dynamical calculation (preserving the flavor symmetries) will have
to respect this condition. In Ref.~\cite{BBS}, using some dynamical
calculation, it is argued that this seems to be the case.

\section{Linear Realizations}

\label{linear}

{}For the linearly realized effective Lagrangian we can start directly with
the triplet field $V_a$. This complex field encodes the five Goldstones plus
a massive scalar. The relation with the non linearly realized fields is
provided in Eq.~(\ref{lvnl}).

While in the non linear realization case the effective Lagrangian is ordered
in number of derivatives, usually in the linear case, the terms are ordered
according to their increasing mass dimension. The general dimension four
effective Lagrangian is:
\begin{equation}
{\cal L}=D^{\mu }V^{\dagger }D_{\mu }V+P(V)\ .
\end{equation}
$P(V)$ is a potential term of the general form:
\begin{equation}
P(V)=-M_{V}^{2}V^{\dagger }V+\lambda _{V}\left[ V^{\dagger }V\right] ^{2}\ ,
\end{equation}
with $M_{V}$ and $\lambda _{V}$ real parameters and the negative
sign mass term has been chosen to provide a non zero vev for $V$

The color covariant derivative is defined as follows:
\begin{equation}
D_{\mu }V=\partial _{\mu }V+iV\,G_{\mu }\ ,
\end{equation}
and $G_{\mu }=G_{\mu }^{m}T^{m}$ is the gluon field. In
constructing the previous lagrangian, we have also assumed parity
invariance. The potential leads to a non vanishing vacuum
expectation value for $V$ which we choose to align in the color
direction $3$,
\begin{equation}
\left\langle V_{a}\right\rangle =\sqrt{2}v\delta _{a}^{3}\ ,
\qquad v=\frac{1}{2}\sqrt{\frac{M^2_V}{\lambda_V}} \ .
\end{equation}
The local $SU(3)$ gauge invariance is then spontaneously broken to
the $SU(2)$ local gauge subgroup. $5$ of the $6$ independent
scalars contained in $V_{a}$ are massless. We choose the unitary
gauge and absorbe the massless degrees of freedom in the
longitudinal components of the 5 massive gauge fields. In this
gauge we can write the fluctuations around the vacuum for $V_{a}$
as:
\begin{equation}
V=\left(
\begin{array}{c}
0 \\
0 \\
\sqrt{2}v+\sigma
\end{array}
\right) .
\end{equation}
where the scalar $\sigma $ is a massive real field with $\langle \sigma
\rangle =0$. The masses of the $5$ gluons are related to the condensate via:
\begin{equation}
m_{{i}}=v,\qquad m_{{8}}=\frac{2}{\sqrt{3}}v\ ,  \label{lrgm}
\end{equation}
with $i=4,5,6,7$. Clearly no Goldstone boson survives since there is no
global symmetry left unbroken. As expected the massless degrees of freedom
are the first $3$ gluons for the unbroken $SU(2)$ gauge symmetry. However
the latter are supposed to confine and hence to generate a new confining
scale associated with pure gluon-dynamics with two colors but no flavors.

We immediately notice, as in reference \cite{R,ssh}, that the linearly
realized effective Lagrangian at the dimension four level predicts the
squared gluon mass ratio between the eighth gluon and the other massive ones
to be $4/3$. This result does not agree with the underlying calculations
\cite{R}. In order to resolve the issue in Ref.~\cite{R,ssh}, it was pointed
out that one can include in the linear Lagrangian another two derivative
term of the form:
\begin{equation}
\frac{b}{v^{2}}\left[ V^{\dagger }D_{\mu }V\right] \left[ V^{\dagger }D^{\mu
}V\right] \ .
\end{equation}
$b$ is a real number. This term has mass dimension 6. At this point we
cannot use, in general, the standard naive dimensional counting arguments
for suppressing other higher dimensional terms for the effective linearly
realized Lagrangian. Interestingly, in the non linear realization case, all
the terms needed to correctly describe the gluon masses arises at the same
derivative order. This fact suggests that non linear realization description
leads to the correct counting scheme.

\section{Conclusions}

\label{conclusions}

We constructed the low energy effective theory describing Quantum
Chromodynamics with two flavors at high density. In order to
correctly implement the low energy theorems, we have used the non
linear realization framework. An important difference with respect
to the 3 flavor case is the presence, guaranteed by 't~Hooft
anomaly conditions, of massless fermions. We have hence properly
added the, in medium, fermions to the effective Lagrangian. Then we
generalized it to the case when Lorentz invariance is broken, by
medium effects, to $O(3)$. We also show that there are two
independent two derivative terms for the Goldstone bosons (which
will eventually become longitudinal components of the color gauge
bosons). This is shown to be an effect related to the baryon number
violation. We finally investigate the linearly realized effective
Lagrangian which, in general, is useful and well defined only when
describing a phase transition. Here we see, as also noted in
Ref.~\cite{R,ssh}, that the naive (in mass) dimensional counting
rule used to naturally order the terms does not hold when
confronted with perturbative dynamical calculations \cite{R}. In
contrast the non linearly realized theory provided a consistent
counting scheme.

\vskip2cm \centerline{\bf Acknowledgments}

It is pleasure for us to thank D.H. Rischke and J. Schechter for
interesting discussions and encouragement. The work of Z.D. and
F.S. has been partially supported by the US DOE under contract
DE-FG-02-92ER-40704.

\end{document}